# Interaction with thermal radiation in the free expansion and mixing of ideal gases and Gibbs' paradox in classical thermodynamics

**A Paglietti**

Department of Structural Engineering, University of Cagliari,

I-09123 Cagliari, Italy

E-mail: paglietti@unica.it

**Abstract.** The standard theory of ideal gases ignores the interaction of the gas particles with the thermal radiation (*photon gas*) that fills the otherwise vacuum space between them. This is an unphysical feature of the theory since every material in this universe, and hence also the particles of a gas, absorbs and radiates thermal energy. The interaction with the thermal radiation that is contained within the volume of the body may be important in gases since the latter, unlike solids and liquids, are capable of undergoing conspicuous volume changes. Taking this interaction into account makes the behaviour of the ideal gases more realistic and removes Gibbs' paradox.

*For an abridged version of this paper see ref.* [9]

## 1. Introduction

A rigid vessel is composed of two equal chambers separated by a removable partition. The two chambers are filled with two different ideal gases at the same initial temperature $T_0$. Let $n_1$ and $n_2$ be the number of moles of gas contained in the first and in the second chamber, respectively. The gases in the vessel cannot exchange heat with the surroundings as the vessel walls are adiabatic. When the partition is removed, which can be done without performing any work on the gases, the two gases diffuse through each other. This is a spontaneous irreversible process and we are interested in determining the entropy increase it produces.

The problem is a standard one in Classical Thermodynamics. Its solution can be found in many textbooks. In particular, the excellent treatise by Fast [1, pp. 41-43] contains a clear approach to this problem, entirely within the realm of Classical Thermodynamics. Being ideal, the two gases do not interact with each other during the mixing process. This means, in particular, that as the partition is removed each of them expands into the volume occupied by the other, as if it expanded in a vacuum. Furthermore, since the free expansion of an ideal gas in a vacuum does not produce any temperature change, the final temperature of the gas mixture must be the same as the initial temperature $T_0$. The entropy change $\Delta S$ due to the considered expansion process is obtained directly from the classical expression of the entropy of an ideal gas:

$$S = n \ (c_v \ln T + R \ln v) + \text{const.} \tag{1}$$

Here $c_v$ is the molar specific heat at constant volume, $n$ the number of moles of the gas, $R$ the universal gas constant, $T$ the absolute temperature, while $v = V/n$ is the volume $V$ per mole. It is thus found that

$$\Delta S = R \ (n_1 \ln \frac{V_1 + V_2}{V_1} + n_2 \ln \frac{V_1 + V_2}{V_2} ). \tag{2}$$

where $V_1$ and $V_2$ are the volumes of the two chambers. In the present case we have $V_1 = V_2 = V$, which simplifies eq. (2) to:

$$\Delta S = R \ (n_1 + n_2) \ln 2. \tag{3}$$

The so-called *Gibbs' paradox* arises because the above formulae are independent of the physicochemical nature of the two gases. This seems hardly acceptable. Surely, the entropy change resulting from mixing together say one mole of helium and one mole of ammonia should be different than that resulting from mixing two similar amounts of two different isotopes of oxygen. In other words a change in the properties of the mixing gases should produce a change in the entropy increase due to the mixing, which is denied by the above formulae.

The same formulae apply in particular when $n_1 = n_2 = 1$. In this case eq. (3) simplifies to

$$\Delta S = 2\,R\,\ln 2 > 0. \qquad\qquad (4)$$

Now, if the two ideal gases are the same, the gas pressure will be the same in the two chambers because they both contain one mole of the gas at the same volume and the same temperature. In this case too the theory predicts the same entropy change (4) following the partition removal. No entropy increase should however occur, simply because no change in the state of the two gases takes place as the partition is removed. This inconsistency in itself is often referred to as the Gibbs' paradox. Actually, it is a further consequence of the fact that eqs (2) and (3) do not depend on the properties of the mixing gases.

The existence of Gibbs' paradox casts some shadows on the classical theory of ideal gases. It suggests that that theory may somehow be flawed even in the range of pressures and temperatures where it turns out to provide an otherwise superb approximation of the behaviour of the real gases.

The same paradox is met when the theory of ideal gases is approached by the methods of Statistical Mechanics. In that case the situation is aggravated by the fact that the traditional expression of the entropy of a perfect gas, as introduced in many textbooks on Statistical Mechanics, contrasts with the notion that entropy should be an extensive quantity. This aspect of the paradox is resolved by introducing the notion of indistinguishability of the various molecules of the gas. This modifies the number of available microstates and, thus, the entropy function (cf., e.g., [2]). Once such a remedy is taken, the discontinuous jump to zero of the entropy change due to the mixing of two gases as the difference in their properties tends to zero, is neither surprising nor paradoxical [2], [3]. It remains the unpalatable fact, though, that when it comes to two different gases, the statistical mechanics approach predicts that the entropy change due to their mixing should not depend on the nature of the gases. A way out of this shortcoming has been proposed by Lin. For a recent commentary of his proposal and reference to the original papers the reader is referred to [4]. Lin's approach, however, introduces the concept of information entropy, which is foreign to Classical Thermodynamics.

In looking for a solution to this problem, one cannot help but observe that Classical Thermodynamics and Statistical Mechanics are essentially two different approaches, resting on entirely different bases. Each approach should therefore be consistent within its own framework. It is only a matter of scientific rigour then, that we should not use one approach to justify the other. For this reason, in the following sections we shall attempt to resolve the various aspects of Gibbs' paradox within the realm of Classical Thermodynamics.

More precisely, we shall show that the origin of the paradox lies in the fact that the notion of ideal gas, as introduced in Classical Thermodynamics or, for that matter, also in Statistical Mechanics, cannot be entirely realistic. This is so because it represents a material that does not radiate any energy. Every material in this universe radiates energy, depending on its temperature. This fact is ignored by the theory of the ideal gases. The latter are imagined as being made of volumeless particles, endowed with mass, unable to interact at distance with each other and with the radiation that always fills the very space in which they are moving about. As a matter of fact, that space will exchange energy with them through thermal radiation until thermal equilibrium is reached. Taking into account this phenomenon will make the ideal gas more physical and also remove Gibbs' paradox.

A similar interaction can be neglected in liquids and solids, since they undergo only minor volume changes.

A comprehensive bibliography on this topic, including more than a hundred papers can be found in [8]. It should be observed, however, that none of the papers quoted there approaches the problem within the framework of classical macroscopic thermodynamics adopted in the present paper.

## 2. The photon gas that fills the vacuum amid the gas particles

Any cavity or otherwise empty space harbours electromagnetic radiation within it. This radiation is often referred to as *thermal radiation* or *photon gas*. The latter terminology is particularly appropriate because from the macroscopic standpoint such a radiation behaves in many ways as a gas. In this section the main formulae concerning the photon gas are recalled, as can be found in many books of Classical Thermodynamics (see e.g. [1, pp.158-162] or [5, Sect. 13.16]). They apply in thermal equilibrium conditions, that is when the walls of the cavity and/or any material particle within it have reached the same equilibrium temperature. The latter will be referred to as the temperature of the photon gas itself. Its value in the absolute temperature scale will be denoted by $T$, as usual.

In thermal equilibrium conditions, a unit volume of space filled with a photon gas stores the energy $u_r$ given by

$$u_r = a T^4 . \tag{5}$$

The constant $a$ appearing here can be expressed as

$$a = \frac{4}{c} \sigma , \tag{6}$$

where $c$ is the velocity of light while $\sigma$ is the well-known Stefan-Boltzman constant [$\sigma$= 56.697 nW/m$^2$ deg$^4$, which makes $a$=75.646 10$^{-8}$ nJ/m$^3$deg$^4$ since $c$=2.998 10$^8$ m/sec].

Eq. (5) shows that the energy density of the photon gas depends entirely on temperature. It does not change, in particular, as the volume of the cavity containing the photon gas is progressively contracted or expanded. In a gas, whether ideal or not, a similar process produces a change in the gas pressure and hence in the gas energy density, because the number of molecules of the gas remains constant as its volume is changed. In contrast, the number of photons of a photon gas in thermal equilibrium in a cavity changes as the volume of the cavity expands or shrinks. This feature is widely known to be one of the main differences between a gas and a photon gas.

From eq. (5) it follows that in thermal equilibrium conditions the internal energy $U_r$ of a volume $V$ of photon gas is given by

$$U_r = a T^4 V \, . \qquad (7)$$

In the same conditions, the photon gas exerts a pressure $P$ on the walls of the cavity, the value of which is

$$P = \frac{1}{3} u_r = \frac{1}{3} a T^4 \, . \qquad (8)$$

This pressure does work as the volume of the cavity expands or contracts. The work supplied by the photon gas to the cavity walls as its volume is increased by d$V$ turns out to be

$$\mathrm{d}W_{out} = P \, \mathrm{d}V = \frac{1}{3} u \, \mathrm{d}V = \frac{1}{3} a T^4 \, \mathrm{d}V \, . \qquad (9)$$

Thus if $Q$ denotes the amount of heat absorbed by the photon gas, we can apply the forst law of thermodynamics to state that

$$\mathrm{d}U_r = \mathrm{d}Q - \mathrm{d}W_{out} \, . \qquad (10)$$

From this and from eqs (7) and (8) we get

$$\mathrm{d}Q = 4 \, a T^3 V \, \mathrm{d}T + \frac{4}{3} a T^4 \, \mathrm{d}V \, . \qquad (11)$$

By applying this equation to a reversible process we obtain

$$\mathrm{d}S_r = 4\,aT^2V\,\mathrm{d}T + \frac{4}{3}aT^3\,\mathrm{d}V\,, \tag{12}$$

since $\mathrm{d}Q = T\mathrm{d}S_r$ for reversible processes. The total differential equation (12) can easily be integrated to give the following expression for the entropy $S$ of the photon gas:

$$S_r = \frac{4}{3}aT^3V\,, \tag{13}$$

the integration constant being set equal to zero, since $S_r = 0$ for $T = 0$.

Finally, from eq. (11) the specific (per unit volume) thermal capacity $r_v$ of the photon gas at constant volume is immediately obtained:

$$r_v = \frac{1}{V}\frac{\mathrm{d}Q}{\mathrm{d}T} = 4\,aT^3\,. \tag{14}$$

## 3. The influence of thermal radiation on the adiabatic free expansion of an ideal gas in a perfect vacuum

According to classical theory, a free adiabatic expansion of an ideal gas should leave its temperature unaltered. This follows from the fact that the particles of the gas do not exert long-range interactions between each other, which means that they can only store kinetic energy. In a free expansion no external work is done by the gas, since it expands in a vacuum. No thermal energy is exchanged either, if the expansion is adiabatic. In these conditions, the kinetic energy of the gas particles is conserved, which in particular means that the gas temperature in the final equilibrium state after the expansion must be the same as the initial one.

The above explanation does not take into account that no material is physically admissible if it does not emit and absorb thermal radiation. The ideal gas is lacking in this respect, which may lead to some physical inconsistencies in its behaviour. In the present section we shall show how the interaction with the ubiquitous photon gas, which cannot be avoided by any real gas –not even in the ranges of temperatures and pressures where its state equations coincide with those of an ideal gas– makes the gas cool as a result of a free adiabatic expansion.

We shall consider here the case in which the gas, though harbouring thermal radiation within its own volume, expands in a space that is empty both of matter and radiation. Such a vacuum will be referred to as *perfect vacuum*. It is of interest in several cosmological problems. The more common case of an ideal gas expanding in a ordinary vacuum, empty of matter but containing radiation (photon gas), will be tackled in the next section.

Let us first of all determine the amount of heat that a photon gas must absorb from the surroundings as it expands at constant temperature. If $\Delta V$ is the volume increase of the gas, the amount of heat needed to expand it isothermally is given by

$$\Delta Q = \frac{4}{3} a T^4 \Delta V \ . \tag{15}$$

This immediately follows from eq. (11) once we set d$T$=0 and integrate the resulting equation between the initial volume $V$ and the final one $V + \Delta V$.

If this amount of heat is not supplied, which is the case when the expansion takes place adiabatically, then the volume change $\Delta V$ will produce a reduction in the photon gas temperature. On the other hand, if the cavity walls are adiabatic but the cavity itself also contains a gas, then the amount of heat $\Delta Q$ will be taken from the particles of the gas, which will cool down as a result. In this case, the temperature change can to a good approximation be calculated by assuming that the thermal capacity of the photon gas is negligible with respect to that of the ideal gas within the cavity. Keeping in mind that the molar specific heat $c_v$ of an ideal gas does not depend on volume and referring to the case in which the cavity contains just one mole of ideal gas, the temperature variation due to the expansion of the latter will be given by

$$\Delta T = -\frac{\Delta Q}{c_v} = -\frac{4}{3} \ \frac{a T^4 \Delta V}{c_v} \quad . \tag{16}$$

If there are $n$ moles of gas in the cavity, this temperature variation should be divided by $n$. In writing eq. (16) we assumed that $\Delta T$ was small enough as to produce a negligible change in $c_v$ during the expansion. This is certainly so in the vast majority of cases. In any case, taking account of the dependence of $c_v$ on $T$ does not appear to pose any serious problem. Of course this temperature change makes the internal energy of the gas change by the same amount $\Delta Q$, since the internal energy of an ideal gas has the well-known expression

$$U_g = c_v \, T + C, \tag{17}$$

$C$ being an arbitrary constant.

The entropy change of the system due to the considered expansion is the sum of the contribution $\Delta S_g$ coming from the ideal gas and the contribution $\Delta S_r$ of the photon gas. The former is readily obtained from eq. (2) and is given by

$$\Delta S_g = c_v \, \ln \frac{T + \Delta T}{T} + R \, \ln \frac{V + \Delta V}{V} \qquad (18)$$

for each mole of gas. The other contribution is calculated from eq. (12) to be given by:

$$\Delta S_r = 4 \, a \, T^2 \, V \, \Delta T + \frac{4}{3} a \, T^3 \, \Delta V \, . \qquad (19)$$

If the volume $V$ is not too large, $\Delta S_r$ can be neglected with respect to $\Delta S_g$. Moreover, for small values of $\Delta T$ as the one involved in the present case, the first term in the right-hand side of eq. (19) can be neglected too, as long as $T$ is sufficiently far from zero. Under these conditions we can, to a good approximation, calculate the total entropy change of the system as

$$\Delta S = R \, \ln \frac{V + \Delta V}{V} \, . \qquad (20)$$

In conclusion, we can say that as far as the considered process is concerned the entropy change of the system is not essentially affected by the presence of the photon gas. The latter, however, does produce a cooling effect in the ideal gas temperature resulting after the expansion. The measure of this effect is given by eq. (16). It shows that $\Delta T$ is proportional to $\Delta V$. This temperature change, therefore, is not a priori negligible, since there is no limit to the extent of the expansion that a gas can suffer.

## 4. The case of an adiabatic expansion in a vacuum with thermal radiation

In many practical cases the cavity where the gas expands is free from matter but filled with thermal radiation. We may speak then of a vacuum with thermal radiation. The presence of the latter in the space where the gas expands exerts a pressure $P$ on the front of the expanding gas. An expansion $\Delta V$ of the gas will therefore make it spend the amount of work $\Delta W_{out}$ given by

$$\Delta W_{out} = P \, \Delta V \, . \qquad (21)$$

I should be observed, in passing, that two ideal gases do not exert any action to each other. For this reason they can expand through each other in the same volume, while ignoring the presence of the other gas. The same is not true when an ideal gas expands in a space filled with photon gas. First of all a photon gas is not an ideal gas, as its state equations (7) and (8) show unequivocally. Secondly and more important, the photon gas does interact with the ideal gas. The reason is that, as already observed, there is no material that does not absorb and emit thermal radiation. So the particles of the ideal gas must absorb and emit energy in the form of thermal radiation and this makes them sensitive to the radiation pressure coming from the photon gas.

Let us then focus our attention to the important case of an ideal gas expanding adiabatically in an otherwise empty cavity containing thermal radiation at the same temperature $T_0$ as the initial temperature of the expanding gas. The expansion process is assumed to take place so slowly that the pressure and temperature of the expanding medium is uniform throughout the process and so is the temperature of the photon gas in the space where the ideal gas expands. In order to calculate the temperature change $\Delta T$ brought about by an expansion $\Delta V$ of the gas, we observe that in the expansion the expanding medium (ideal gas plus photon gas) absorbs the heat

$$\Delta Q_\mathrm{s} = a\, T^4 \Delta V \tag{22}$$

that comes from the photon gas in the empty part of the vessel as the volume of the latter is reduced by $\Delta V$. By applying the energy balance law to the expanding medium we therefore get:

$$\Delta U_\mathrm{g} + \Delta U_\mathrm{r} = -\Delta W_\mathrm{out} + \Delta Q_\mathrm{s} \,. \tag{23}$$

Here $\Delta U_\mathrm{g}$ denotes the internal energy change of the ideal gas, $\Delta U_\mathrm{r}$ the energy change of the photon gas within the volume of the expanding gas, $\Delta W_\mathrm{out}$ is given by eq. (21) and represents the work done by the expanding medium against the pressure $P$ exerted by the outside photon gas. In view of eq. (8) this work can be expressed as

$$\Delta W_\mathrm{out} = \frac{1}{3}\, a\, T^4\, \Delta V \,, \tag{24}$$

while from eq. (17) we obtain

$$\Delta U_{\mathrm{g}} = c_{\mathrm{v}} \, \Delta T, \tag{25}$$

where

$$\Delta T = T_1 - T_0 \tag{26}$$

Finally, the quantity $\Delta U_{\mathrm{r}}$ can be obtained directly as the difference between the value of $U_{\mathrm{r}}$ calculated for $T_0 + \Delta T$ and $V + \Delta V$ and the value of the same quantity calculated for $T_0$ and $V$. From eq. (7) to thus obtain:

$$\Delta U_{\mathrm{r}} = a \, (T_0 + \Delta T)^4 \, (V + \Delta V) - a \, T_0^{\,4} \, V \; . \tag{27}$$

Since the temperature change $\Delta T$ is expected to be very small, we can approximate the first term in the right-hand side of this equation by a Taylor expansion about $T_0$. We thus get

$$\Delta U_{\mathrm{r}} = a \; T^4 \Delta V + 4 a \, T^{\,3} \, (V + \Delta V) \Delta T \; , \tag{28}$$

where we set $T$ for $T_0$, since the formula applies to any value of the initial temperature $T_0$.

By introducing eqs (28), (25) and (24) into eq. (23) we obtain:

$$\Delta T = -\frac{1}{3} \; \frac{a \, T^4 \Delta V}{c_{\mathrm{v}} + 4 \, a \, T^{\,3} \, (V + \Delta V)} \quad . \tag{29}$$

Unless the volume $V + \Delta V$ is very large, the term $4 \, a \, T^{\,3} \, (V + \Delta V)$ can be neglected with respect to $c_{\mathrm{v}}$, on the account that $a$ is a very small quantity. With this approximation, eq. (29) reduces to

$$\Delta T = -\frac{1}{3} \; \frac{a \, T^4 \Delta V}{c_{\mathrm{v}}} \quad . \tag{30}$$

## 5. The entropy change of the two gases ensuing from the partition removal

Let us now refer to the two gases filling the two separate chambers of the adiabatic vessel considered in Sect. 1. When the partition is removed, the two gasses expand by flowing one through the other. As already recalled, each gas expands as if it were in a vacuum, since the two gases are supposed to be ideal. This enables us to calculate the entropy change of the system of the two gases by referring to the following sequence of processes:

1   A free adiabatic expansion of each separate gas, bringing them to the same final volume $2V$. In this process, the gas does not supply/absorb any work to/from the surroundings since the process can be assimilated to a free expansion in a vacuum. As observed in Sec. 3, the process will produce a change in the entropy and temperature of each expanding gas, though. Moreover, different gases will suffer different temperature changes.

2   A transfer of heat between the two expanded gases bringing them to the same final temperature T*. Being a heat transfer from a hotter gas to a colder one, this process will result in a further entropy change of the system. In the process, no heat is exchanged with the surroundings as the vessel walls are adiabatic.

3   Finally, the two gases are allowed to mix reversibly at the constant volume $2V$ they reached at the end of phase (1). A process of this kind is admissible from the physical standpoint and was first conceived by Plank [6, Sect. 236] (see also [1, p. 41] and [3, p. 729]). Being adiabatic and reversible, the process will not produce any entropy change. It will not produce any change in the temperature of the gases either, since the latter are ideal and the process does not bring about any volume change. The system will thus reach the final state in which the two gases are mixed together at temperature T* and volume 2V.

The details of each of the above three phases of the entire process are worked out below.

**Phase 1.** (*Free adiabatic expansion*)

One mole of an ideal gas fills the volume $V$ of one chamber of the vessel. Let $T_0$ be its initial temperature and $c_v'$ its molar specific heat at constant volume. The gas expands adiabatically in a vacuum filled with radiation at temperature $T_0$, until it occupies the volume $2V$ of the entire vessel, thus increasing its volume by $\Delta V = V$. If the presence of heat radiation is taken into account, the equilibrium temperature $T_1$ that results after the expansion can be calculated from eq. (30) to be given by:

$$T_1 = T_0 - \frac{1}{3} \frac{a T_0^{\,4} \, \Delta V}{c_v'} \quad . \tag{31}$$

A similar adiabatic expansion of the gas filling the other chamber will bring it to the final volume $2V$ and to a final equilibrium temperature $T_2$ given by:

$$T_2 = T_0 - \frac{1}{3} \frac{a T_0^{\,4} \, \Delta V}{c_v''} \quad , \tag{32}$$

where $c''_v$ is the molar specific heat at constant volume of the gas in the second chamber. The above temperature changes are obtained from the approximate eq. (30). The gist of the arguments that follow, however, remain valid if the temperature changes are obtained from the more rigorous eq. (29).

In order to calculate the entropy change brought about by this phase of the process, we first observe that if $\Delta V$ is not too large, both $T_1$ and $T_2$ will be sufficiently near to $T_0$ as to allow us to apply the approximate equation (20). This process will therefore increase the entropy of each gas by the amount

$$\Delta_e S = R \ln 2. \tag{33}$$

This will produce an overall increase in the entropy of the two gases by the amount

$$\Delta_1 S = 2 \Delta_e S = 2 R \ln 2. \tag{34}$$

It is apparent that such an entropy change coincides with that which would be expected for an ideal gas in the absence of the photon gas. The presence of the latter, however, will produce a cooling effect that is proportional to $\Delta V$ resulting from the considered expansion, as predicted by eqs (31)-(32). Small as this effect may be (it is proportional to $\Delta V$, though!), it cannot be avoided not even by an ideal gas, since every material must absorb and radiate thermal energy. Most importantly, the amount of cooling depends on the specific heat of the gas, so that it will generally be different for different ideal gases.

**Phase 2.** (*Internal heat transfer*)

The previous process brings the two gases to states $(T_1, 2V)$ and $(T_2, 2V)$, respectively. To be definite, we shall assume that $c'_v > c''_v$, which in view of eqs (31) and (32) implies that $T_1 > T_2$. We now put the two gases in thermal contact with each other while keeping them thermally insulated from the surroundings. In these conditions the hotter gas will supply heat to the colder one until both gases reach the same equilibrium temperature $T^*$. The determination of $T^*$ is an elementary problem of heat transfer. It is solved by equating the total amount of heat $Q_1^*$ lost by the hotter gas to the amount of heat $Q_2^*$ absorbed by the colder one. Since we are considering one mole of each gas, we have

$$Q_1^* = (T_1 - T^*) \, c'_v \tag{35}$$

and

$$Q_2^* = (T^* - T_2) \, c_v'' .$$ (36)

We therefore get

$$T^* = \frac{T_1 \, c_v' + T_2 \, c_v''}{c_v' + c_v''} .$$ (37)

Once the amounts of heat (35) and (36) are known, the entropy change caused by the process can also be obtained. Since the temperatures $T_1$, $T_2$ and $T^*$ are close together, we shall not introduce any serious mistake if we assume that all the heat $Q_1^*$ is lost at the constant temperature $T_1$ and that all the heat $Q_2^*$ is absorbed at constant temperature $T_2$. The entropy change due to the heat transfer is, accordingly:

$$\Delta_2 S \; = \; -\frac{Q_1^*}{T_1} + \frac{Q_2^*}{T_2} \; = \; \frac{c_v' \, c_v''}{c_v' + c_v''} \; \frac{(T_1 - T_2)^2}{T_1 \, T_2} ,$$ (38)

which is clearly greater than zero. The important point to be noted here is that this entropy change depends on the specific heats of the mixing gases. Such dependence is both explicit, as shown by equation (38), and implicit through $T_1$ and $T_2$ via equations (31) and (32). In the case in which $c_v' = c_v''$, the latter equations yield $T_1 = T_2$, which makes $\Delta_2 S = 0$. In particular, $\Delta_2 S$ vanishes if the two chambers are filled with the same gas.

**Phase 3.** (*Isothermal reversible mixing*)

As previously observed, in this phase of the process the two perfect gases do not suffer any change in entropy or temperature since they undergo a reversible isothermal mixing. The presence of the photon gas does not alter this conclusion, since this phase of the process takes place at a constant volume.

From eqs (34) and (38) the entropy change due to the whole process turns out to be, therefore:

$$\Delta S = \Delta_1 S + \Delta_2 S \; = \; 2 \, R \ln 2 + \frac{c_v' \, c_v''}{c_v' + c_v''} \; \frac{(T_1 - T_2)^2}{T_1 \, T_2} .$$ (39)

The first term in the far right-hand side of this equation is due to the volume expansion and has the same value no matter the ideal gases under consideration. The last term attains different values for different mixing gases.

Incidentally, such a result helps to define precisely when two ideal gases are to be treated as the same or as different as far as the thermodynamics of their mixing is concerned. It shows that it all depends on whether they have the same specific heat or not. Other differences in their properties are not relevant as far as the thermodynamics of the process is concerned. This fact should be taken into due account when a Statistical Mechanics approach to this phenomenon is sought. It implies that some otherwise different microstates of a gas mixture should be considered to be the same if they are relevant to two different ideal gases which possess the same specific heat.

## 6. The solution of the paradox

According to the analysis of the previous section, there are two contributions to the entropy change $\Delta S$ that follows the removal of the partition between the two ideal gases contained in the adiabatic vessel we considered in the Introduction. The first contribution is $\Delta_1 S$ and is relevant to what we called Phase 1 of the process. This entropy change in this phase arises from the expansion of each gas separately, and is practically independent of their particular physicochemical properties [cf. eq. (33)]. The final temperature resulting from the process, however, *does* depend on the properties of the expanding gas and, more precisely, on its specific heat. This means that this part of the process will in general produce different final temperatures for different expanding gases [cf. eqs. (31) and (32)].

The other contribution to the entropy change, namely $\Delta_2 S$, results from Phase 2 of the process. At a variance with $\Delta_1 S$, this entropy change depends on the specific heats $c'_v$ and $c''_v$, of the two gases. Moreover, as the difference between $c'_v$ and $c''_v$ tends to zero, so does $\Delta_2 S$. As already observed, this shows that as far as Phase 2 is concerned, the difference in the specific heats of the two gases can be taken as a measure of their difference. Any property of the gases other than their specific heats has no effect on the entropy change resulting from this part of the process. Such a result provides a solution to Gibbs' paradox, which is entirely within the framework of classical macroscopic thermodynamics. The arguments in support to this claim are discussed below.

For clarity's sake, in what follows we shall decompose Gibbs' paradox into the following three paradoxes A to C, and give a separate solution to each of them.

**Paradox A:** "*The entropy of mixing is independent of the nature of the two gases*"

**Solution**. Since $\Delta S = \Delta_1 S + \Delta_2 S$, from the above results we can conclude that the total entropy change upon mixing depends on the difference in the specific heat of the mixing gases. This resolves this part of Gibbs' paradox, which objected to having the same entropy change no matter the physicochemical properties of the two gases. The paradox arises if the effect of thermal radiation is neglected. The established formulae also show that small variations in the difference of the specific heats of the two gases produce small variations in the predicted value of $\Delta S$, which complete the answer to this part of the paradox.

**Paradox B:** "*According to eq. (4), the entropy change resulting from the mutual expansion of the two gases during their mixing is different from zero even if the two gases are the same*"

**Solution**. Consider the particular case in which the gases in the two chambers are the same, at the same initial pressure and temperature. In this case the two gases will not expand at all as the partition is removed. This is so because, after the partition removal, the number of gas particles that go from chamber 1 to chamber 2 will be equal to the number of gas particles that go from chamber 2 to chamber 1, the two gases being in thermal equilibrium with each other. Since the particles of the two gases are identical, the net effect of this particle exchange will be the same as if the partition was not removed at all. In the absence of any expansion, the final volume of each gas in each of the two chambers remains the same as it was before the partition removal. Eqs (2) to (4) do not therefore apply to this case. The general equation (1) is of course still valid. From that equation it follows that $\Delta S=0$ when temperature and volume are constant, which is quite correct. In the present case, therefore, the theory predicts that the partition removal should not produce any entropy change, thus solving the paradox. Observe that no consideration about thermal radiation is needed to answer this aspect of Gibbs' paradox.

**Paradox C:** "*The slightest difference in the physicochemical properties of the two mixing gases produces a finite entropy change as they mix together*"

**Solution**. In the case of two identical gases, the entropy change due to mixing vanishes, while the slightest difference in their specific heat would produce a non-vanishing finite entropy change $\Delta S = \Delta_1 S + \Delta_2 S$. There is no inconsistency here, though. The point is that the volumetric contribution $\Delta_1 S$ always applies in the case of two different gases −no mater how little their specific heats differ from each other, as long as they are not the same. As observed above, however, the contribution $\Delta_1 S$ is missing in the case of two identical gases, since no expansion takes place in that case. It should be obvious then that one should not compare the entropy change due to a process that includes an expansion, with that of another process that does not. It should also be observed that if we confine our attention to the contribution $\Delta_2 S$, which is *not* due to the expansion, then the entropy change relevant to the case of two different gases will tend smoothly to zero as $c'_\mathrm{v}$ and $c''_\mathrm{v}$ tend to a same value $c_\mathrm{v}$. This is perfectly consistent with the fact that $\Delta S=0$ in the limit case in which the two gases are identical and answers the same-gas-part of Gibbs' paradox concerning the alleged entropy jump inconsistency.

**Remark**. It might still be objected that if we put a red gas in chamber 1 and a white gas in chamber 2 we would get a pink mixture once the partition is removed, even if the two gases are "*the same*" in that $c'_\mathrm{v} = c''_\mathrm{v}$. This could be interpreted as the evidence that the entropy of the system increases, contrary to what we have just concluded above. Such a change in colour, however, is not a thermodynamic process. "Redness" and "whiteness" are not state variables of the system. Nor do they enter the state equations of the gases. As a consequence, no internal energy or entropy change can be produced in the system by a colour change of the gas. No work or heat is absorbed in the process either. The gas colour change that would certainly take place in the considered situation is a purely mechanical process; it is a consequence of the disordered distribution of the velocities of the gas particles. It could also be regarded as a demonstration of the thermal agitation of the gas particles, which is always active in any gas, even in thermal equilibrium conditions, provided that $T\neq0$.

But, is the pink state of the two gases more disordered than the initial red and white one? Perhaps. In some sense at least. However, as remarked in [7], disorder and macroscopic entropy are not always related.

## 7. Conclusions

A common feature of every known material in this universe is that it absorbs and emits thermal radiation. The ideal gas cannot be an exception. By adopting an approach that is entirely confined within the framework of classical macroscopic thermodynamics, we have shown that, if proper account of the thermal equilibrium between an ideal gas and its own thermal radiation is taken, the free expansion of the gas cannot be isothermal. The resulting temperature change is proportional to the volume change, which in gases can be as large as one may like. It also depends on the specific heat of the gas itself. Beside of being of interest in itself, the latter feature eliminates the well-known inconsistencies that in the classical macroscopic thermodynamics are associated to the mixing of two ideal gases and are collectively referred to as Gibbs' paradox.